\documentclass{sigchi-ext}
\usepackage[T1]{fontenc}
\usepackage{textcomp}
\usepackage[scaled=.92]{helvet} %
\usepackage{graphicx} %
\usepackage{balance}  %
\usepackage{booktabs} %
\usepackage{ccicons}  %
\usepackage{ragged2e} %

\usepackage{xspace}

\usepackage{marginnote} 
\def\plaintitle{Modeling Behaviour to Predict User State: Self-Reports as Ground Truth} 
\def\plainauthor{Julian Frommel, Regan Mandryk}

\def\plainkeywords{emotion recognition; user state; supervised learning; machine learning; questionnaire; ground truth}

\title{\plaintitle}

\numberofauthors{2}

\author{%
    \alignauthor{%
        \textbf{Julian Frommel}\\
            \affaddr{University of Saskatchewan, Saskatoon, SK, Canada} \\
            \email{julian.frommel@usask.ca}
    }
    \alignauthor{%
        \textbf{Regan L. Mandryk}\\
        \affaddr{University of Saskatchewan, Saskatoon, SK, Canada} \\
        \email{regan@cs.usask.ca}
    }
}

\definecolor{linkColor}{RGB}{6,125,233}
\hypersetup{%
  pdftitle={\plaintitle},
  pdfauthor={\plainauthor},
  pdfkeywords={\plainkeywords},
  bookmarksnumbered,
  pdfstartview={FitH},
  colorlinks,
  citecolor=black,
  filecolor=black,
  linkcolor=black,
  urlcolor=linkColor,
  breaklinks=true,
}

\begin{document}

\CopyrightYear{2020}
\copyrightinfo{Position Paper at the Momentary Emotion Elicitation \& Capture (MEEC) Workshop at ACM CHI 2020}

\maketitle

\RaggedRight{} 

\begin{abstract}
Methods that detect user states such as emotions are useful for interactive systems. In this position paper, we argue for model-based approaches that are trained on user behaviour and self-reported user state as ground truths. In an application context, they record behaviour, extract relevant features, and use the models to predict user states. We describe how this approach can be implemented and discuss its benefits in comparison to solely self-reports in an application and to models of behaviour without the self-report ground truths. Finally, we discuss shortcomings of this approach by considering its drawbacks and limitations. 

\end{abstract}



\section{Introduction}
Capturing complex user states such as emotions can be valuable for interactive systems. Systems that can detect emotional state and react appropriately can improve the users' performance and satisfaction~\cite{epp2011}, decrease negatively valenced states such as frustration~\cite{klein2002computer, picard1999affective}, enhance player experience in games~\cite{10.1145/3242671.3242682}, and improve learning outcomes in serious games~\cite{Schrader2017}. %
To leverage these benefits, robust and powerful methods to detect user state are necessary. Self-report methods such as questionnaires are hindered in applications due to interruptions of user experience~\cite{10.1145/2793107.2793130} and imprecision if referring to events that happened too long ago~\cite{mauss2009measures}. In addition, they cannot capture changes in states when only used retrospectively once after an experience. Methods using behavioural features are limited by a lack of appraisal and hard to interpret by researchers and developers as there is no common language translation of state representations. %

In this position paper, we argue for methods employing user behaviour and self-reports in combination. In particular, we propose the value of methods using models trained on user behaviour to predict self-reported user state, e.g., questionnaires responses, effectively using them as ground truth. In earlier work in a gaming context, we used this approach to create models of player behaviour predicting self-reported valence, arousal, and dominance~\cite{10.1145/3242671.3242672}, and self-reported affiliation between co-players~\cite{frommel_recognizing_2020}. Similarly, other work has trained models of user behaviour to predict self-reported experiences such as perceived difficulty, aesthetics, and enjoyment of game levels~\cite{summerville2017understanding}, and emotional state based on keystrokes~\cite{epp2011}. While we draw on examples from gaming, this approach can be used in a wide variety of application contexts, such as mobile apps, productive work settings, automotive contexts, or mixed reality setups. %

This approach has benefits over using self-reports alone; it does not interrupt the interaction and experience in the application context, it increases temporal resolution over post-experience questionnaires, and it decreases required human input because of an increased automation. On the other hand, the combination of behaviour and self-reports with models has benefits over pure behaviour-based modeling, e.g., with a strong foundation in theory and a high interpretability by developers and researchers. In this position paper, we will describe the technical approach and discuss its benefits and limitations.

\section{Technical Approach}
This approach usually involves supervised machine learning techniques, as described by the following steps:
\begin{enumerate}
    \item \textbf{Definition:} First, researchers define the user states of interest, the self-report measures used to assess them, and the indicators to use for prediction, e.g., user behaviour features. 
    \item \textbf{Data Collection:} In a training phase, participants similar to the end users will engage in interaction with the application. This training phase can be a pre-study or a pre-experience phase such as a tutorial if it is similar to the end user experience. During this interaction, they use self-reports of user states while their behaviour is recorded. This can involve a wide variety of self-report measures such as repeated single-item responses (e.g., Likert scales such as ``Currently I feel frustrated.``), more complex post-experience questionnaires, or even think aloud protocols. 
    \item \textbf{Data Preparation:} Researchers prepare data for training. This involves cleaning, extraction of relevant features based on domain expert knowledge, and slicing of continuous data. By creating time windows spanning between self-reports, it is possible to assign particular phases of user behaviour to the experienced user states. 
    \item \textbf{Training:} Then, models can be trained in a supervised learning approach that uses behavioural features as input and self-reported user state as output. 
    \item \textbf{Prediction:} In the application context, user behaviour is recorded and fed to the model that predicts user state, e.g., the most likely emotion in a classification approach or the level of continuous arousal via regression. 
    \item \textbf{Refinement:} It is possible to refine the models by continuously collecting behavioural data, analyzing the users' reactions, and occasional checks for user state via self-reports.
\end{enumerate}
Models trained with multiple users are generalizable to a certain degree and can be particularly potent when the specific user is already known to them. With this approach, behaviour is used to predict self-reported user state. 

\section{Advantages}
This method has a variety of benefits.

\textbf{Unobtrusiveness:}
In comparison to employing self-reports during the interaction, this approach can be considered unobtrusive in a way that it does not interrupt the experience or affect the users. Ideally, the measurement is so unobtrusive that users can forget that their state is measured. By moving the interruptions through self-reports to a pre-experience training phase, it is enough to record the user's behaviour, extract features, and use them to predict states. This way, this approach can be used in a wide variety of applications, in which interruptions would negatively affect the experience.

\textbf{Temporal Resolution:}
We propose models that are trained on time windows of user behaviour, for which users provide self-reports of experience. With this approach, there are shorter periods of user behaviour that relate to a particular state, e.g., for individual gestures~\cite{10.1145/3242671.3242672}. In contrast to traditional post-experience questionnaires, each individual time window can be analyzed and used to predict user state, effectively increasing the temporal resolution of capturing. As such, the methods are better suited to detect user states that change dynamically over time.

\textbf{Computational Detection of Feature Importance and Differentiation:}
With user behaviour, it can be challenging to assess which indicators are important and how they relate to user states. A computational approach that uses machine learning to create models can be useful in this regard. Interactions amongst features and their relationship to outcomes can be detected with supervised learning techniques.  %
As such, they are well-suited to analyze the relevance  of behavioural features and the decision criteria of assigning them to outcome states. %
As such, using self-reports as ground truth can beneficial for user state assessment because of a machine learning's ability to create complex models.

\textbf{Context Relevance:}
There is a huge variety in user states and it is very challenging to train methods that can capture the state suitable for all contexts. For instance, a model can capture the users' emotion by differentiating between the six basic emotions as defined by Ekman~\cite{ekman1992argument}, but then might not be well suited to capture curiosity. As such, it makes sense that methods are trained for states that are context-relevant and to employ ground truths that are relevant for the specific application context. Self-reports lend themselves for this, as models are trained specifically with data from training phases similar to the application context. %

\textbf{Interpretability:}
Learning models that predict questionnaire responses can be beneficial because they provide a ground truth that is easily interpretable by designers, developers, and researchers. Behaviour-based methods are sometimes limited when they are not used in combination with self-report measures, but only to differentiate between conditions. In this case, it is only possible to detect that behaviour differs, but the direction of effect is not always obvious and thresholds are not necessarily understandable for researchers. Physiological signals that are used without self-reports are challenging to interpret, e.g., at which threshold is a heart rate reflective of a fun or meaningful experience? Self-report measures can provide this additional context that is easily understandable for researchers.

\textbf{Variability in Outcomes:}
With system variations, it is possible to elicit different states. For instance, in our work~\cite{10.1145/3242671.3242672}, we used two variants of a game to elicit a wider range of emotional responses. In addition to the regular game, we used a version with manipulated feedback to generate more negatively valenced emotional responses to gather a dataset with more variance in outcomes. This way, our method was trained with data comprised of positive and negative emotions that was elicited through manipulation. A major advantage of using self-reports as ground truths is their ability to act as immediate validation. They can be used to verify that the manipulation elicits the intended states.

\paragraph{Grounding in Theory:}
Theoretical foundation is highly important in research~\cite{rogers2012hci}. Self-report measures are particularly pertinent to provide a theoretical grounding for user state capturing methods. There is a myriad of validated scales and questionnaires that measure different aspects of user state. Frequently, they build on theoretical models for the particular experience or state that they measure. As such, a method that predicts self-report responses on such a questionnaire builds on these models as well, e.g., by considering subcomponents of a state by taking into account the questionnaire's subscales.

\paragraph{Grounding in User Perception:}
Self-reports as ground truth are beneficial because they are grounded in an user's appraisal of a situation. As such, they provide information about how they perceived an experience, which is beneficial if a particular scenario can be interpreted differently. In the context of emotion, for example, a joke can be perceived as funny by some, leading to joy, while it might trigger sadness in others because it reminds them of personal tragic events. As such, a user's interpretation can be necessary for researchers to understand user experience.

\paragraph{Personalization:}
Model-based approaches are well-suited for repeated usage~\cite{10.1016/j.intcom.2005.05.002}. The training phase could be leveraged to train models that know the individual characteristics of different users. For instance, digital games frequently feature tutorials that teach game mechanics. One could imagine a scenario where pre-trained models are refined with the data from players in these tutorials. They would answer short self-reports of their experience during the tutorial while their behaviour is recorded. This would be used to adjust the models with the data of the individual players to create models that can leverage the information of the individual characteristics of a specific player, e.g., a particular way they react when they are frustrated.

\paragraph{Applicability to Complex States:}
There is increasing interest in the prediction of highly complex states such as affiliation~\cite{frommel_recognizing_2020} or mental health~\cite{info:doi/10.2196/13485}. 
Using self-report measures as ground truth facilitates the application of assessment approaches for such states by providing a ground truth that is challenging to gather otherwise. Self-reports generally lend themselves as a first source of information for assessment and can be useful to assess such states in early studies.

\section{Disadvantages}

Drawbacks and limitations can impede applicability.

\subsection{Drawbacks}

\textbf{User Data Collection:}
To create models that predict self-reports, it is necessary to collect user data, e.g., with pre-studies, in which users answers questionnaires. While powerful, this warrants some effort. Researchers should be aware that they have to collect enough user data to train valid models that ideally can generalize beyond individual users. 

\textbf{Privacy:}
The analysis of user behaviour can affect their privacy, e.g., with video or audio features. As such, researchers have to consider if it is worth it to employ potentially invasive recording methods to assess user state, which can be problematic if it affects aspects that are not relevant to the application context, such as persons in the same room as the user. 

\subsection{Limitations}

\textbf{Reliance on Ground Truth Validity:}
The validity of this approach strongly relies on the validity of the ground truth measurements. The user state assessment is only valid if the self-reports actually reflect the user's state. As such, our method is limited in the same way as its ground truth of self-reports in general, e.g., by social desirability~\cite{krumpal2013determinants} %
or recall bias~\cite{mauss2009measures}. Therefore, researchers have to pay attention that the self-reports measure the intended state as desired.

\textbf{Generalizability to Application Context:}
The models might be biased because they are trained in a context where users know that they are tested, which can affect their behaviour~\cite{webb1999unobtrusive}. As such, they might behave somewhat differently in the training phase if they know that their behaviour is analyzed in comparison to how they would behave otherwise. However, it makes sense to assume that their behaviour is still useful to predict their self-reported experiences, because users would know that their state is analyzed in an application scenario as they have to consent to the use of analysis methods.

\textbf{Interruptions:}
Gathering self-report measures during training effectively still leads to interruptions. This is especially problematic with lengthy questionnaires. %
However, interrupting the experience during training is preferable to interrupting the end user experience. As such, researchers should be wary of these interruptions, deliberate if they affect their models, and preferably use short self-report measures that are less disruptive. In the end, we argue that the benefits of increased temporal resolution generally outweigh the negative effects of interruptions in training.

\textbf{Behavioural Features:}
To train valid models that employ user behaviour, it is necessary to use behavioural features that are in fact indicative of the outcome states to predict. Not every behavioural trace can be used to predict user state, e.g., if there is no generalizable connection between them. As such, researchers should consider which features are good indicators for what they aim to predict and use these as input to their models. Previous research can inform the selection of such features by suggesting potentially valuable features.

\section{Predicting Self-Reported Affective State}
In the context of emotion recognition and capture, this approach lends itself to predicting self-reported affect based on different theoretical models of affect. There is a wide variety on theoretical models of affect with multiple widely used constructs~\cite{ekkekakis2012affect} and many self-reports measures that have been proposed to measure them. By using models of user behaviour to predict self-reported affect, it is possible leverage these validated scales. As such, it is possible to use questionnaires that measure dimensional conceptualizations of affect based on valence, arousal, and dominance (e.g., with the \emph{Self-Assessment Manikin, SAM}~\cite{bradley1994measuring}), mood (e.g., \emph{Positive and Negative Affect Schedule, PANAS}~\cite{watson1988development}), or specific emotional states (e.g., with the \emph{Positive and Negative Affect Schedule - Expanded Form, PANAS-X}, or the \emph{Discrete Emotions Questionnaire, DEQ}~\cite{harmon2016discrete}~\cite{watson1999panas}). Therefore, these approaches can be applied for the specific conceptualization that is important for a researcher's use case. Similarly, a wide variety of behavioural features can be considered that has been shown to be connected to affect, such as physiological reactions~\cite{cowley2016psychophysiology, doi:10.1080/01449290500331156} or interaction parameters~\cite{epp2011, 10.1145/3242671.3242672}. It is further possible to use different questionnaires in the training phase to train multiple models that use behaviour to predict the user state based on different theoretical models (e.g., SAM and basic emotions). If models are trained anyways, additional questionnaires can be used in the training phase to create models that predict different aspects of user state. As such, using behavior to predict self-reported user state can be beneficial for researchers who are interested in the users' emotional states.

\section{Conclusion}
In this position paper, we discussed benefits and limitations of user state assessment approaches that use supervised learning to train models of behavioural features to predict self-reported state. Researchers should be wary of the disadvantages and use this model-based approach if its benefits outweigh its limitations. In particular, it is important that self-report measures have to be available, researchers can conduct a pre-study to collect training data, and that the limitations of self-reports do not prohibit the application. If these limitations are less important in a particular context, researchers should consider model-based approaches with user behaviour predicting self-reports of user state and experience, e.g., their affective state. With this approach, researchers can use models to predict user states, such as emotions, with an approach that has an increased temporal resolution, and is powerful, widely applicable, and easy to interpret.

\section{Acknowledgements}
We thank NSERC and SWaGUR for funding.

\balance{} 

\bibliographystyle{SIGCHI-Reference-Format}
\bibliography{extended_abstract}

\end{document}